\documentclass[a4paper,12pt]{article}
\usepackage[english]{babel}
\usepackage[a4paper,tmargin=3truecm,bmargin=3truecm,rmargin=2.5truecm,
lmargin=2.5truecm,twoside,verbose=true]{geometry}

\usepackage{cancel,graphicx}
\usepackage[dvips]{hyperref}

\usepackage{amsmath,amssymb}

\usepackage{amsmath,amssymb,slashed}
\usepackage[amsmath, hyperref, thmmarks]{ntheorem}
\numberwithin{equation}{section}

\numberwithin{equation}{section}

\newcommand{\be}{\begin{equation}} 
\newcommand{\ee}{\end{equation}} 
\newcommand{\bea}{\begin{eqnarray}} 
\newcommand{\eea}{\end{eqnarray}} 
\newcommand{\edo}{\end{document}}

\newcommand {\rd} {{\,\rm d}}

\usepackage{graphicx}

\theoremsymbol{}
\theorembodyfont{\slshape}
\theoremheaderfont{\normalfont\bfseries}
\theoremseparator{}
\newtheorem{Theorem}{Theorem}[section]
\newtheorem{theorem}[Theorem]{Theorem}

\newtheorem{lemma}[Theorem]{Lemma}

\allowdisplaybreaks[1]

\renewenvironment{thebibliography}[1]
         {\section*{References}\frenchspacing\small
          \begin{list}{[\arabic{enumi}]}
         {\usecounter{enumi}\parsep=2pt\topsep 0pt
         \settowidth{\labelwidth}{[#1]}
         \leftmargin=\labelwidth\advance\leftmargin\labelsep
         \rightmargin=0pt\itemsep=1pt\sloppy}}{\end{list}}

\title{Using mixed data in the Inverse Scattering Problem}
\author{M. Lassaut$^{a,b}$, S.Y. Larsen$^c$, 
S.A. Sofianos$^b$, J.-C. Wallet$^d$}

\begin{document}

\maketitle
\vspace*{-1cm}

\begin{center}

\textit{$^a$Institut de Physique Nucl\'eaire\\
CNRS/Universit\'e Paris-Sud 11 (UMR8608)\\
F-91406 Orsay Cedex, France\\
    e-mail: \texttt{lassaut@ipno.in2p3.fr}}\\
\textit{$^b$Physics Department, University of South Africa,\\
   Pretoria 0003, South Africa\\
    e-mail: \texttt{sofiasa@science.unisa.ac.za}}\\
\textit{$^c$Department of Physics, Temple University,\\
    Philadelphia, PA 19122, USA\\
    e-mail: \texttt{syl@temple.edu}}\\
\textit{$^d$Laboratoire de Physique Th\'eorique, B\^at.\ 210\\
    CNRS, Universit\'e Paris Sud-11,  F-91405 Orsay Cedex, France\\
    e-mail: \texttt{jean-christophe.wallet@th.u-psud.fr}}\\
\end{center}
\vskip 2cm

\begin{abstract}
Consider the fixed-$\ell$ inverse scattering problem.
We show that the zeros of the regular solution of the Schr\"odinger 
equation, $r_{n}(E)$, which are monotonic functions of the energy, 
determine a unique potential when the domain of the energy is such 
that the $r_{n}(E)$ range from zero to infinity.
This suggests that the use of the mixed data
of phase-shifts 
 $\{ \delta(\ell_0,k), k \geq k_0 \} \cup
\{ \delta(\ell,k_0), \ell \geq \ell_0 \}$,
for which the zeros of the regular solution are monotonic in both 
domains, and range from zero to infinity,
offers the possibility of determining the potential 
in a unique way. 
\end{abstract}

\maketitle

\pagebreak

\section{Introduction}

Approaches to the three-dimensional inverse scattering problem can be classified in two categories\cite{newt,CS}.
 In the first case, the fixed-$E$ problem, Loeffel\cite{Loef68} 
obtained theorems predicting a unique potential from 
the knowledge of the phase-shifts  $\delta(\ell,k)$ at a specific energy 
$E=k^2$, for all (non-discrete) non negative values of $\lambda=\ell+1/2$.
When the set of data is reduced to discrete values of $\lambda=\ell+1/2$, $\ell$ non-negative integers,
Carlson's theorem\cite{Carlson}  predicts a unique potential   $V(E,r)$,  
provided that this latter belongs to a suitable class\cite{CS,Loef68}. 
The  Newton series 
allow the construction of the  potential 
$V(E,r)$ from this latter  set of discrete values\cite{Loef68}.

In the second case, known also as  fixed-$\ell$ 
problem,  (see Ref.\cite{CS} and references therein), an $E$-independent
potential   $V_\ell(r)$ satisfying the integrability conditions 
\be
    \int_b^{+\infty} \vert V_\ell(r) \vert \rd r  <  \infty\,, \qquad 
            \int_0^{+\infty} r  \vert V_\ell(r) \vert \rd r  <  \infty\,,
 \qquad\quad b>0\,,
\label{int}
\ee
can be  constructed from the phase-shifts $\delta(\ell,k)$, given for all 
momenta $k\in (0,\infty)$ and from the discrete spectrum data
(eigen-energies and the corresponding normalization constants).

 Another class of Bargmann potentials can be constructed from
 input data which  are partly $E$-dependent and partly
$\ell$-dependent and where  $a E + b \ell (\ell+1)$ is a 
constant\cite{RZ}. This latter construction is based on an extension of 
Newton's method. 

Historically, the idea of mixing the data was first explored 
by Grosse and Martin\cite{GM}  for confining potentials.
They conjectured that the knowledge of the ground state energies 
$E_{\ell}^{(0)}$, for all non-negative integers $\ell$, 
allows the recovery of the potential in an unique way.
This problem has been studied numerically in Ref.\cite{YIL}.

In the present work, we are also concerned with a non-standard inverse problem, 
namely with the construction of the potential from a spectrum which involves  
both data coming from the $E$-fixed problem and the $\ell$-fixed problem, 
but where extensions of Newton's method can no longer be applied.
The set of mixed  data considered here  is 
$\{ \delta(\ell_0,k), k \geq k_0 \} \cup
 \{ \delta(\ell,k_0), \ell \geq \ell_0 \}$,  corresponding to the set of scattering parameters  $\{ E \geq E_0, \ell=\ell_0\} \cup
\{ E=E_0, \ell \geq \ell_0  \}$.
 
We will investigate in detail,  for the fixed $\ell$-problem, to what extent the knowledge of 
the zeros $r_n(E)$ of the regular solution allows the determination
of the potential. We will show that  the function $r_n(E)$, being  monotonic 
with respect to  the energy,
reflects  a unique potential, provided that the domain of the energy is 
such that $r_n(E)$ ranges from zero to infinity.
This will lead to a uniqueness theorem. We also will depict a method 
which allows the recovery of the piecewise constant potentials from the 
knowledge of the function $r_n(E)$. 
The advantage is that this latter method can be extended to our non-standard 
spectrum, given above, because the zeros of the regular solution
are monotonic in both domains and still range from zero to infinity.
The uniqueness theorem still applies and the piecewise constant potentials 
can be recovered from the zeros of the regular solution\cite{LLSW}.
    
The next step consists in investigating  to what extent the mixed data 
determines  a unique potential.
By analogy with the fixed $\ell$-problem we expect, as explained in detail 
below,  that this should be the case,  provided  the potential 
satisfes the integrability conditions (\ref{int}).

In Sec.{\bf 2} we discuss the formalism leading to the mixed
inversion scheme presented in Sec.{\bf 3} and, as an example, we apply it
to piecewise constant potentials.
 The uniqueness of the method,    demonstrated in Ref. \cite{LLSW}
by using a JWKB  procedure will then be examined  
in the Born approximation.
In Sec.{\bf 4} we will present our conclusions.

\section{Formalism}
Consider a potential 
$V(r)$ satisfying the integrability conditions\cite{CS} 
\bea
            \int_0^{a} r  \vert V_\ell(r) \vert \rd r  & <  & \infty\,, \qquad\quad   a<\infty \nonumber\\
 \int_b^{+\infty} \vert V_\ell(r) \vert \rd r  & < &  \infty\,, 
 \qquad\quad b>0\,
\label{intp}
\eea
and the Schr\"odinger equation
\be
\left(\frac{\rd^2}{\rd r^2}\right. +\left. 
   E-V(r) -\frac{(\ell + 1/2)^{2}-1/4}{r^2} \right)   \psi_{\ell}(E,r) =  0\,.
\label{Schr}
\ee
The regular solutions $\psi_{\ell}(E,r)$
satisfy the Cauchy condition
$\lim_{r \to 0} \psi_{\ell}(E,r) r^{-\ell -1}=1$ and take on the
asymptotic form
$
      \psi_{\ell} \propto \sin(k r - \ell \pi/2 + \delta(\ell,k))
$
 with $ \delta(\ell,k)$ being  the phase shifts. (Recall that here $E=k^2$. )

 Let us define  $r(\ell,E)$ such that the regular solution 
 $\psi_{\ell}(E,r)$ vanishes for $r=r(\ell,E) \neq 0$.
For a fixed energy 
and a fixed $\ell$, there is then  a countable number of zeros,
each position being denoted by  $r_n(\ell,E)$.  

Furthermore, for any potential,
the zeros satisfy the monotonicity   properties:
\begin{itemize}
\item   $ E=k^2 \mapsto r_n(\ell,E)$ is a decreasing function, as  has been shown by Sturm 
  in 1830's\cite{Sturm} and,
\item  \ $\ell \mapsto r_n(\ell,E)$
is an increasing function. This has been  shown in Ref.\cite{LLSW}. 
\end{itemize}

Recall that  for potentials satisfying  (\ref{intp}) the function $r_n(\ell,E)$ is such that $r_n(\ell,E)\to 0$ for $E\to\infty$\cite{Chadan67}.
In the absence of bound states  
$ r_n(\ell,E)\to\infty$ for $E\to0$. 
However, in the presence of one bound state, 
$ r_1(\ell,E)\to\infty$ for  $E\to E_1$. Here $E_1 < 0$
denotes the 
ground state energy\footnote{We adopt this notation to denote 
the ground state by $E_1$ instead of the traditionally used $E_0$
in order to be consistent with the meaning given to $n$, namely 
to denote zeros of the wave function}. 

For $N$ bound states,  $E_1 < E_2 ... < E_N$, if $n \leq N $, 
$ r_n(\ell,E)\to\infty$ for  $E\to E_n$
\cite{Chadan67}, 
whereas for $n > N$, $ r_n(\ell,E)\to\infty$ as  $E\to0$.
\begin{figure}
\centering
\hspace*{-1.cm}
\includegraphics[width=8.5cm, height=6.cm]{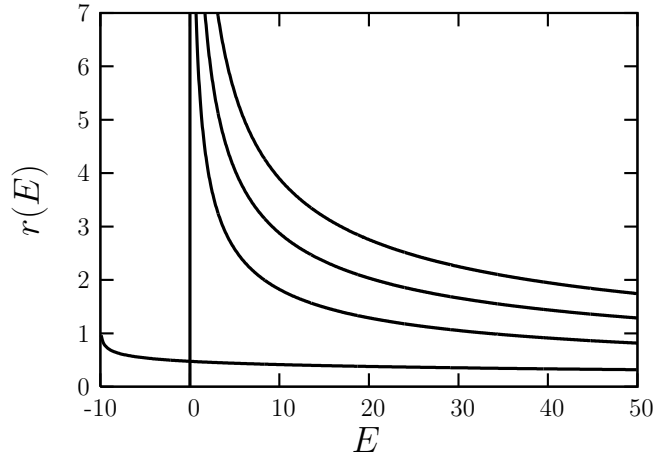}
\caption{\label{f1}
Zeros $r_n(E), n=1,2,3$ and $4$, of the regular 
s-wave solution for a Bargmann potential  (see text).}
\end{figure}
%

This is illustrated in  figure {\bf 1} where we  have drawn the four first 
zeros of the regular  $s$-wave solution 
for the Bargmann transparent potential, Eq.(27) of  Ref. \cite{LLSR}  with $\ell=0$ and $\gamma^2=10$, which has 
a bound state at the energy $E=-10$ in $1/L^2$ units.
The vertical axis $E=0$ is explicitly depicted to show that all zeros but the first one go to infinity as $E$ goes to $0$.
The first zero goes to infinity when $E \to E_1=-10$.

In the fixed-$\ell$ inversion, if
all  the zeros are assumed to be known, i.e. $E \mapsto r_n( \ell,E)$ is
known and $r_n(\ell,E)$ describes the entire interval $[0,\infty[$, 
the potential $V_\ell(r)$ is then uniquely determined.
This can be checked easily. Consider the following Sturm-Liouville 
problem on $[0,R]$, i.e., namely  the equation

\be
    \psi_{\ell}''(r)+\left(E-V_{\ell}(R-r) -\frac{\ell(\ell+1)}{(R-r)^2} \right)  \psi_{\ell}(r)=0,
\label{stur}
\ee
coupled with the Dirichlet conditions
$$\psi_{\ell}(0)=\psi_{\ell}(R)=0 \ .$$ 

The spectral data are the eigenvalues, namely the $E_n^*$'s such 
that $R=r_n(\ell,E_n^*)$, together with the normalization 
constants\cite{Sacks}
$$\rho_n=\frac{\int_0^R \psi_{\ell}(r')^2 dr'}{\psi'_{\ell}(0)^2} \ , $$
here given by the positive values 
$$\rho_n=-(dr_n/dE)(\ell,E_n^*) \ .$$

The potential, assumed to be square integrable on $[0,R]$ is uniquely determined on $[0,R]$ by 
the spectral data $\{ E_n^*,\rho_n\} $\cite{GL}.  
The technique of constructing $V$ from this set of data  is 
well-known\cite{CS,Sacks,CM,CKM}.

 We now concentrate on the case of the zeros of the regular solution.
For a fixed $\ell$, the nth zero of the
regular solution can be considered as a function of the energy $r_n(\ell,E)$,
monotonic and such that $\psi_{\ell}(E,r_n(\ell,E)) \equiv 0$.
It defines a line of zeros. Moreover  $E \mapsto r_n(\ell,E)$ 
 admits an inverse function $r \mapsto E_{n,\ell}(r)$ which is also monotonic and is the inverse of this line of zeros.
For example consider the potential $V \equiv 0$  in the s-wave. The regular solution is proportional to 
$\sin(\sqrt{E} r )$. The lines of zeros are given by $r_n(0,E)=n \pi/\sqrt{E}$ and the inverses of these lines are $E_{n,0}=n^2 \pi^2/r^2$.

 We have shown above that if we know all lines   
$r \mapsto E_{n,\ell}(r), n \geq 1$,
for $r$ running from $0$ to $\infty$, then we can construct the desired 
potential,
assumed to be  locally $L^2({\cal R})$.  
If  only a single
line $r \mapsto E_{n_1,\ell}(r)$ is known, no method is available to
recover the potential, except in the special case of piecewise constant potentials.
What we can show is a uniqueness property, if 
$E_{n_1,\ell}(r)$ is known for all positive $r$.

In the special case of piecewise constant potentials, having discontinuities
at values of $r = a_j$, $j =1,\ldots ,j_{max}$, and being zero for
$r > a_{j_{max}}$, it is easy to show that there is a one to one 
correspondence  between the discontinuities of the third
derivative of $E_{n_1,\ell}(r)$ with respect to $r$ 
and  the discontinuities of $V$.
This suggests the following lemma:\par 

\begin{lemma}
 For a piecewise constant potential, the knowledge of a single line of
 zeros allows the reconstruction of
the potential in an unique way provided that the line is 
monotonic with respect to the energy and  runs over the entire positive axis.
\end{lemma}

 The proof is based on  the consideration of the third derivative of 
$$          \psi(E(r),r) \equiv 0 , $$
 where the function  $r \mapsto E(r)$ denotes the inverse  function of the function
$E \mapsto r(E)$ which is  the single line of zeros $r(E)$ considered.
To be specific, if $E'''$ has a
discontinuity at $r=a$  then

\be
        \frac{\rd^3 E}{\rd r^3}(a^+)-\frac{\rd^3 E}{\rd r^3}(a^-)
         =-2 \frac{\rd E}{\rd r}(a)
           \left [V(a^+)-V(a^-)\right] \ .
\label{disc1}
\ee

This is equivalent to the relation
\be
      \frac{\rd^3 r}{\rd E^3}(E_a^+)-\frac{\rd^3 r}{\rd E^3}(E_a^-)  
           =2\left(\frac{\rd r}{\rd E}(E_a)\right)^3 
          \left[V(r(E_a^+))-V(r(E_a^-))\right]  \ .
\label{disc2}
\ee
Since the potential is zero for $r> a_{j_{max}}$, the relation (\ref{disc1}) allows us
to reconstruct the potential between $a_{j_{max}}$ and $a_{j_{max}-1}$. The
procedure can be repeated at each $a_j$, and the potential is obtained at
$r \neq a_j$  by summing the successive values at each
discontinuity appearing beyond $r$.

As an illustration consider for instance the explicit potential defined by 
\bea
V(r) & = &  -2  \qquad\quad r < 2 \nonumber\\
V(r) & = & -1 \qquad\quad 2 < r < 3 \\
V(r) & = & 0 \qquad\quad r>3  \nonumber
\label{pcpot}
\eea
In the figure {\bf 2}, we have drawn the function 
$r \mapsto -E'''(r)/(2 E'(r))$, related to the inverse $E(r)$ of the first 
line of zeros for the $s$-wave regular solution  of the Schr\"odinger 
equation involving the potential (\ref{pcpot}).
\begin{figure}[th]
\centering
\hspace*{-1.cm}
\includegraphics[width=8.5cm, height=6cm]{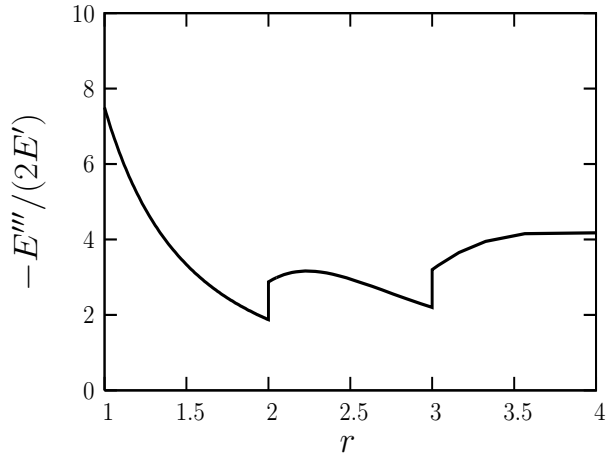}
\caption{\label{f2}
Function $-E'''(r)/(2 E(r))$ (see text).}
\end{figure}
%

 Clearly the discontinuities of  
 $r \mapsto -E'''(r)/(2 E'(r))$ happen at the points 
 where $V$ has discontinuities,  namely $r=2$ and $r=3$,
  and  equation (\ref{disc1}) is satisfied. 
  We know that the potential is zero beyond $r=3$.
  So we have $0=V(3^+)$. From the curve $V(3^+)=V(3^-)+1$ then $V(3^-)=-1$.
As $V$ is piecewise constant we have $V(3^-)=V(2^+)$ and from the curve 
$V(2^+)=V(2^-)+1$ so that $V(2^-)=-2$.
We then recover the potential  of Eq.(\ref{pcpot}).

This method cannot be applied to a potential defined by a continuous function.
Nevertheless, for such potentials, the following uniqueness theorem holds:

\begin{theorem} 
{\it Consider  two  potentials $V_1$ and $V_2$ satisfying (\ref{intp}) and 
locally constant in the vicinity of zero. Within the fixed-$\ell$
  problem, assume that two integers $n_1$ and $n_2$ exist
 such that the $n_1$-th line of zeros for the regular solution for $V_1$ 
 coincides with the $n_2$-th line of  zeros for the regular solution for $V_2$,
  both lines describing the whole positive axis as the energy $E$ varies.  
  Then $V_1\equiv V_2$. } \\
\end{theorem}

The proof is not reproduced here  and given  in Ref. \cite{LLSW}.

\section{Mixed Problem}
Consider the set  $\{ E \geq E_0, \ell=\ell_0 \} 
\cup \{ E=E_0,\ell \geq \ell_0 \}$.
The zeros of the regular solution form a line with two parts.
In the first part, the zeros go  from $r=0$ to $r(\ell_0,E_0)$ as
the energy $E$ varies from $\infty$ to $E_0$ ($\ell_0$ being fixed);   
in the second part, they go from 
$\ell_0$ to $\infty$ ( $E_0$ being fixed).
The monotonicity property required in the lemma and the theorem
is preserved on both domains. As it has been done in  the section {\bf 2}  
we can define a line of zeros.
 The nth line of zeros $r_n(\ell,E)$ of the regular solution describes a line 
formed of two parts. It is defined as follows.
 In the first part the zeros $r_n(\ell_0,E)$ go from $r=0$, 
($E=\infty$) to $r_0=r(\ell_0,E_0),(E=E_0)$ as the energy 
$E$ varies from $\infty$ to $E_0 \ (\ell_0 $ being fixed); in the second part, $r_n(\ell,E_0)$ goes from $r_0=r(\ell_0,E_0), (\ell=\ell_0)$
 The inverse function  of the line of zeros are  defined as follows.
For $r \leq r_n(\ell_0,E_0)=r_0$ it is given by $E$ such that $r=r_n(\ell_0,E)$.
For $r \geq r_n(\ell_0,E_0)$ it is given by $\ell$ such that $r=r_n(\ell,E_0)$.
It is continuous at $r=r_0$ with a discontinuous derivative at $r=r_0$.

 In  Ref. \cite{LLSW} we have shown that 
the above lemma is still valid 
for piecewise constant potentials and  that the uniqueness theorem still works.

To summarize the situation,  we have shown in Ref.\cite{LLSW} 
that a single line of zeros,  which, for the data considered,
always goes from zero to infinity and moreover is
monotonic, determines the potential uniquely. 
The remaining question is to
examine whether  the set of  mixed data
\be
      \{ \delta (\ell=\ell_0,k) \ \ k \in[k_0, + \infty[ \} 
          \cup \{ \delta(\ell,k_0) \ \ell \in[\ell_0,+\infty[ \}
\label{specd}
\ee
 associated to the set $\{\ell=\ell_0, k \in [k_0,\infty[ \} 
\cup \{k=k_0, \ell \in [\ell_0,\infty[\}$,
 determines a line of zeros, and thus the potential, in a unique way -
 which is suggested by the analogy with the $\ell$-fixed problem. 
In the absence of bound states,  
all lines of zeros $E \mapsto r_n(\ell,E)$, monotonic with respect to $E$, 
range from zero ($E$ infinite) to infinity ($E=0$) when $E$ goes from 
infinity to zero. 
In this case, we know that the potential,  when it satisfies (\ref{int})
 is recovered in a unique way, given the phase-shifts $\delta(\ell,k)$ for 
all $k \geq 0$.
In contrast to the condition (\ref{intp}), the condition  (\ref{int}) excludes 
all pathologies  i.e. potentials behaving  asymptotically
like $1/r^2$, encountered in particular in the presence of a zero energy bound 
state, or ghost components in the Jost function\cite{LLSR}. 
With the mixed data we are in the same situation, namely all the lines of zeros 
are monotonic and range from zero ($E$ infinite, $\ell=\ell_0$)
 to infinity ($E=E_0$, $\ell$ infinite). 
So we expect that (\ref{specd}) is associated with a 
 unique potential decreasing faster than $1/r^2$ at infinity.
We cannot prove this in the general case, but we have investigated the problem
in a JWKB approach in Ref. \cite{LLSW},  
where we have shown that the mixed data (\ref{specd}) allow us to recover the 
 potential, provided that the turning point is unique.

It is of interest to investigate what happens in the Born approximation. 
In the seventies Reignier\cite{Reignier} used a 
Born approximation of the scattering
amplitude to show that the knowledge of the phase-shift at a fixed energy
say $E_0=k_0^2$ for each integer $\ell$ is equivalent to 
the knowledge of the 
Fourier sine transform of the potential $r V(r)$,
\be
  g(q)= \int_0^{\infty} \sin(q r) \ r V(r) \rd r\,,
\ee
 for $q \leq 2 k_0$. 
The scattering amplitude is determined from the phase-shifts at fixed energy $\delta(\ell,k_0)$ for $\ell=0,1,2,3,...$, $E=E_0=k_0^2$.

Generally, the integral is assumed to be 
zero for $q >2 k_0$\cite{CS,Sabatier} leading 
to potentials 
\be
         r V(r)= \frac{2}{\pi} \ \int_0^{2 k_0} \sin(q r) \,g(q)\, \rd q
\ , \label{b1} \ee

such that $r V(r)$ is an entire function of $r$ of order 1. 
Other extensions of $g(q)$  are studied in   
Ref.  \cite{Sabatier04}. 

A possible way to extend our knowledge of
$g(q)$ beyond $2 k_0$  is to take the Born approximation for the missing
phase shifts $\delta(\ell=0,k)$ for $k \geq  k_0$. This is given by
\be
       \int_0^{\infty} \sin(k r)^2 \ V(r) \rd r =- k \delta(\ell=0,k)
\ . \ee
The derivative with respect to $k$ yields
\be
 g(q)=  \int_0^{\infty} \sin(q r)  \ r V(r) \rd r =- \frac{\rd 
      (k \delta(\ell=0,k))}{\rd k}\,,\qquad
           \forall q=2 k \geq 2 k_0\,.
\ee
This implies that $g(q)$, known for $q \leq 2 k_0$ is now known for every positive $q$, including $q \geq 2 k_0$, and that $V(r)$ is uniquely given by 
\be
         r V(r)= \frac{2}{\pi} \ \int_0^{\infty} \sin(q r) \,g(q)\, \rd q
\ . \label{b2}
\ee

Consequently, the knowledge of 
 $\{\delta(\ell,k_0), \ell \in {\mathcal N} \} \cup 
 \{\delta(\ell=0,k),k \geq  k_0 \}$ 
allows us to recover the potential in Born approximation 
if $k_0$ is sufficiently high. 

\section{Conclusion}

In the present work, 
we have been  concerned with a non-standard inverse problem, 
namely with the construction of the potential from a spectrum 
which involves   data coming from both the $E$-fixed problem 
and the $\ell$-fixed problem, but where extensions of Newton's 
method can no longer be applied. For the $\ell$-fixed problem,
we have investigated to what 
extent the knowledge of the zeros of the regular solution allows 
the determination of the potential, and  have also given a 
uniqueness theorem. Furthermore, we have  shown that the zeros of the 
regular solution of the Schr\"odinger equation, which are 
monotonic functions of the energy, $r_{n}(E)$, determine a unique 
potential when the domain of energy is such that the $r_{n}(E)$ 
range from zero to infinity.   
The knowledge of a single line of zeros does 
not allow us to recover the underlying potential, except in the special 
case of piecewise constant functions.

As an application of the method, we have considered the mixed data
 $\{ \delta(\ell_0,k) \  \ k \geq k_0 \} \cup
\{ \delta(\ell,k_0) \  \ \ell \geq \ell_0 \}$
for which the zeros of the regular solution are monotonic in both 
domains, and range from zero to infinity.
These mixed data offer the possibility of determining the potential 
in a unique way. Indeed we have shown that a single line of zeros underlies a unique potential, which can be extracted when the potential is a piecewise constant function.
We know from Ref.  \cite{LLSW} that  the mixed data yield a unique $\ell$- and $E$-independent potential,  in the JWKB approximation 
and in the case of a single turning point. 

We have shown that, in Born approximation the following mixed data 
\be
      \{\delta (\ell=0,k),   k \in [k_0, + \infty[ \}
            \cup \{ \delta(\ell,k_0), \ell \in 
            {\cal N} \}
\ee
lead to an unique potential, still assumed to be $\ell$- and $E$-independent, 
which is the inverse Fourier sine transform of a function deduced from the data.

 To conclude, our method, which does not use any extension of the Newton's
 method, applies  to mixed set of data and/or  
 to the "generalized eigenvalue problem"  
 namely the scattering problem  where the function 
 which involves the scattering parameter is no longer separable 
 in the scattering parameter 
and the space coordinates as it happens for the 
fixed-$E$ (function $(\ell,r) \mapsto \ell (\ell+1)/r^2$) and fixed-$\ell$ 
(function $(E,r) \mapsto E$) problems.

Given a set of phase-shifts corresponding to a domain where  
the scattering parameter(s) varies 
(vary),   we conjecture 
 that, if  all lines of zeros of the regular solution  are monotonic,
 continuous   and range from zero to
 infinity when the scattering parameter(s) describes the domain 
 considered, both limits $0$ and $\infty$ being reached at the
 boundary of the domain, then there 
exists a unique potential satisfying (\ref{int}) and 
  corresponding to the set of phase-shifts.

\section*{Acknowledgments} 

We are grateful to R.J. Lombard  for many discussions and
a careful reading of the manuscript.
One of us (ML) is very grateful to the University of South Africa
 for the kind hospitality extended to her.


\end{document}